# High-Q mid-infrared chalcogenide glass resonators

Hongtao Lin[1], Lan Li[1], Yi Zou[1], Spencer Novak[2], Kathleen Richardson[2], and Juejun Hu[1]

1. Department of Materials Science and Engineering, University of Delaware, Newark, Delaware 19716

2. College of Optics and Photonics, CREOL, Department of Materials Science and Engineering, University of Central Florida, Orlando, Florida 32816

Abstract:

In this letter, we fabricated and characterized chalcogenide glass resonators monolithically integrated on mid-infrared transparent $CaF_2$ substrates. The devices feature an intrinsic Q-factor of $4 \times 10^5$ at 5.2 µm wavelength, which represents the highest Q-factor reported in planar mid-infrared resonators. It is found that moisture can significantly impact the device performance when the device was exposed to an ambient environment, although the high-Q characteristics can be restored after undergoing an annealing treatment. Using these devices, we further demonstrated on-chip cavity-enhanced spectroscopy to quantify mid-IR absorption of ethanol solutions around 5.2 µm wavelength.

Introduction:

Integrated planar photonics have been recognized as a preferred platform over discrete optical components for data communications, imaging and field-deployed sensing applications as they uniquely combine small footprint, ruggedness, reduced power consumption, and low cost. However, most integrated photonic devices have been demonstrated in the near-infrared telecommunication wavelengths, and components operating in the mid-infrared (mid-IR) spectral domain have been much less explored, despite their immense application potential in sensing, imaging and free-space communications.[1] One technical challenge for photonic integration in the mid-IR is that silicon oxide, arguably the most commonly used cladding material in integrated photonics, has a short IR cut-off wavelength at 4 µm. Alternative cladding materials including sapphire, compound semiconductors, silicon nitride, and $LiNbO_3$ have been used to resolve the issue. Among these materials, halides and chalcogenides such as $CaF_2$, $MgF_2$, and ZnSe boast broadband mid-IR transparency and extreme low optical loss and are thus particularly attractive for mid-IR photonic integration.

In this letter, we describe monolithic integration of high-performance mid-IR resonator devices on $CaF_2$ substrates. Chalcogenide glasses (ChGs), amorphous compounds containing S, Se and/or Te are chosen as the resonator material given their broad mid-IR transparency window, large photothermal figure-of-merit[7], and most importantly, their amorphous structure and low processing temperature which facilitate monolithic integration on virtually any substrates. We have recently exploited this unique "substrate-blind" integration capacity of ChGs to enable novel functional devices including flexible photonic circuits on plastic substrates[8], nanophotonic light trapping on thin c-Si solar cells, and mid-IR waveguide integration with cascade lasers. Here we leverage this property of ChGs to demonstrate a high-performance, versatile glass-on-CaF$_2$ platform for mid-IR integrated photonics. As shown in the infrared transmission spectra in Fig. 1, the glass-on-CaF$_2$ system can potentially operate over a wide wavelength range from 600 nm to 7.5 µm, only limited by the phonon absorption onset in the CaF$_2$ substrate at 7.5 µm.

The ChG micro-disk resonators were fabricated using a negative-resist-based lift-off process[9] adapted to fabrication on CaF$_2$ substrates. Lithographic patterning was performed using standard i-line UV exposure on a mask aligner. Ge$_{23}$Sb$_7$S$_{70}$ was chosen as the glass composition for this application given its superior thermal and chemical stability compared to its classical As-based counterparts (As$_2$S$_3$ or As$_2$Se$_3$). The Ge$_{23}$Sb$_7$S$_{70}$ glass film was deposited using single-source thermal evaporation following protocols described elsewhere[10].Figure 1a shows the SEM cross-section of a Ge$_{23}$Sb$_7$S$_{70}$ glass waveguide. The small cross-section waveguide supports a single quasi-TE mode shown in Fig. 1b. A pulley coupler design was used to increase coupling strength between the feeding waveguide and the micro-disk resonator as shown in Fig. 1d[11].

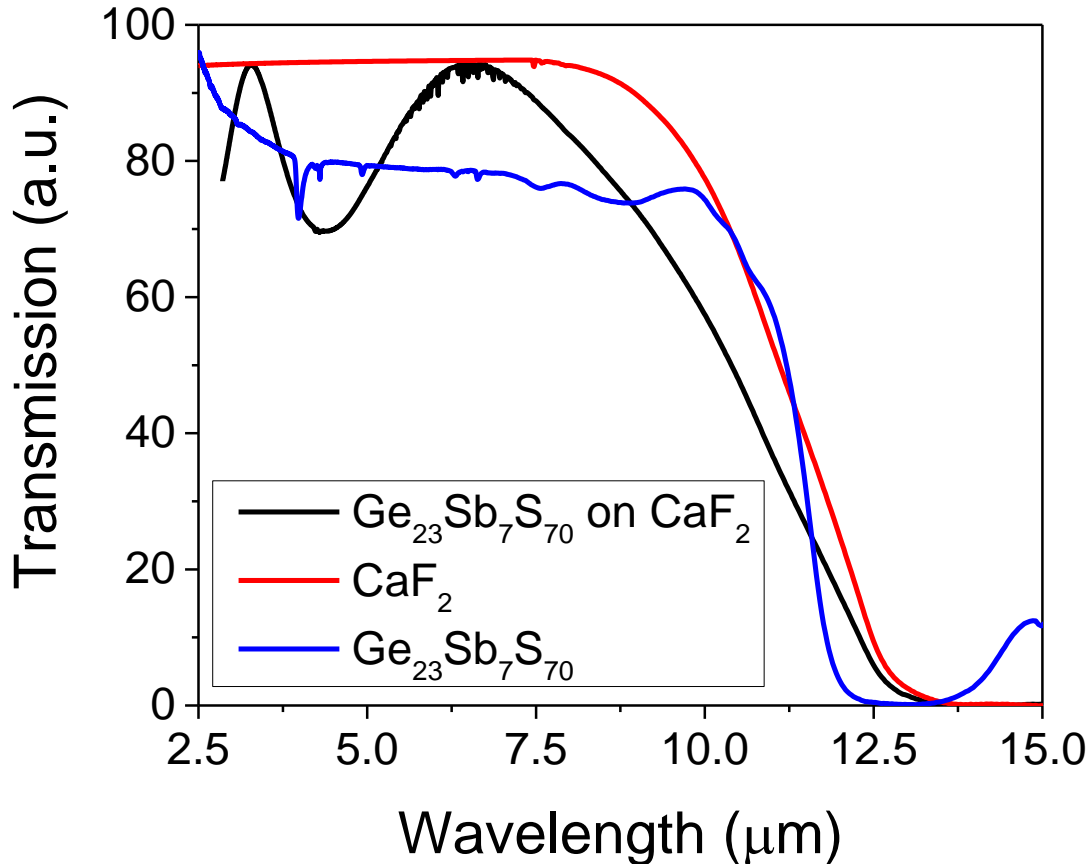

Figure 1. FTIR transmission spectrum of bulk Ge$_{23}$Sb$_7$S$_{70}$ glass, 1 mm thick CaF$_2$ substrate and 1.8 µm thick Ge$_{23}$Sb$_7$S$_{70}$ glass film on CaF$_2$ substrate. The spectra are not corrected for Fresnel loss.

Mid-IR optical transmission characteristics of the devices were measured using a fiber end-fire coupling method around 5.2 µm wavelength. Details of measurement setup can be found elsewhere[12]. Figure 3a plots the transmission spectrum measured from a 100-µm-radius micro-disk resonator. Free spectral range (FSR) of the resonator measured from the spectrum is 20.52 nm. The corresponding group index of the whispering gallery resonant mode is 2.1, which agrees well with our finite-element modal simulations. Notably, doublet peaks were consistently observed on our resonator devices, clearly indicating the lift of degeneracy between two standing wave modes caused by roughness backscattering[13].

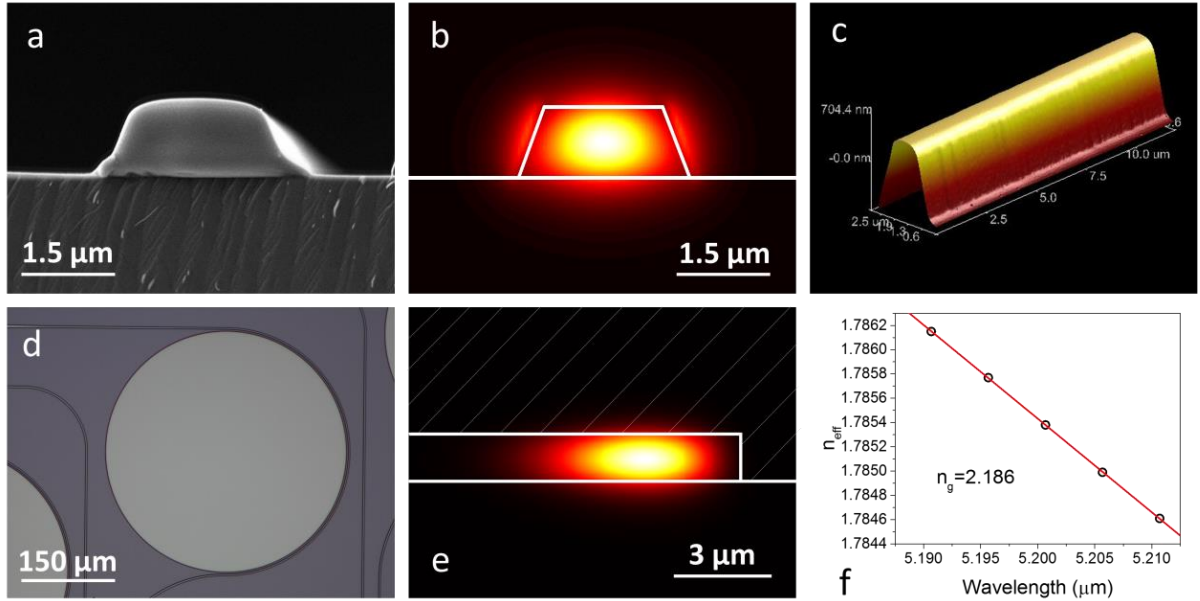

Figure 2. (a) SEM cross-section image and (b) mode profile of a 3 µm by 1.3 µm $Ge_{23}Sb_7S_{70}$ on $CaF_2$ single mode waveguide at 5.2 µm wavelength; (d) microscope image of 100 µm radius pulley coupled disk resonators inset shows the well-defined gap between the bus waveguide and resonator; (e) profile of first order whispery gallery mode of a 100 µm disk resonator.

Transmission spectra of the resonator devices were fitting using the couple mode theory[14] which also takes into account coupling between clockwise and counter-clockwise whispering gallery modes due to backscattering. Fig. 3a shows an exemplary transmission spectrum measured on the device. The Q-factors were extracted by direct fitting of the transmission spectra as well as a method described in supporting information, and both approaches yield identical results. The device exhibits an intrinsic quality factor of ($4 \times 10^5$, which corresponds to a low modal propagation loss of 0.15 dB/cm and an effective optical path length of 29 cm (estimated from $1/\alpha$, where $\alpha$ is the propagation loss in $cm^{-1}$). To the best of our knowledge, this represents the highest Q value ever reported in mid-IR on-chip resonators.

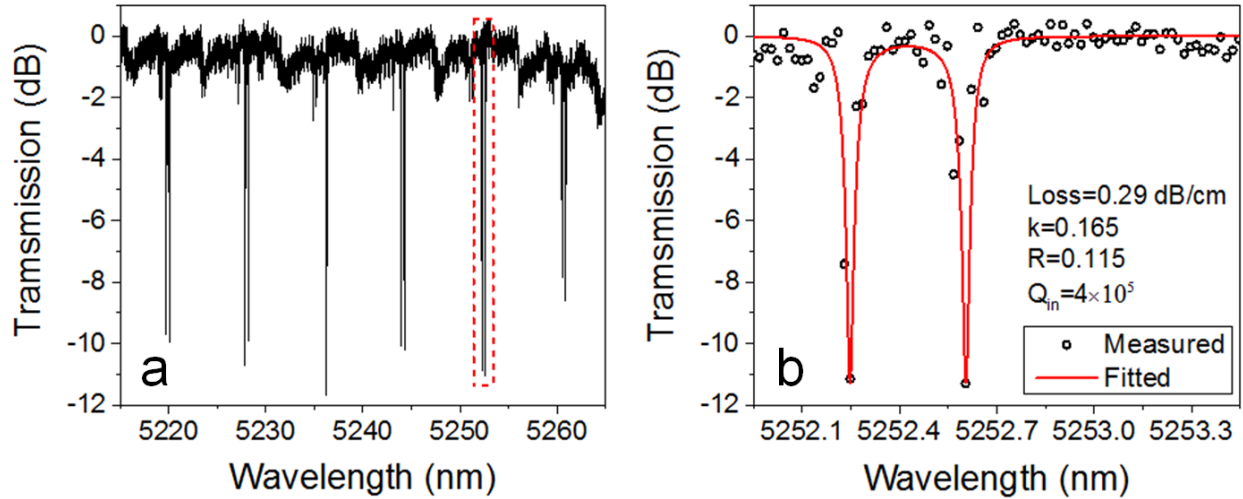

Figure 3. (a) Mid-IR transmission spectrum from a 3 μm width 1 waveguide with 1.75 μm gap pulley coupled disk resonator with 100 μm radius and (b) spectrum near optical resonance at 5202.6 nm wavelength (red box in a). Measured data were fitted by coupled mode theory, and the intrinsic quality factor is about $4 \times 10^5$.

The long effective optical path length in the high-Q resonators also suggest that they are extremely sensitive to changes in their surroundings. As two examples highlighting such sensitivity, we report experimental quantification of the impact of moisture in the ambient environment on the performance of the devices, and proof-of-concept demonstration of cavity-enhanced mid-IR absorption spectroscopic sensing. Figure 3a shows the evolution of a resonator transmission spectrum near the same resonant peak measured over time. While the peak split remained constant in the process, both Q-factor and extinction ratio drastically decreased after exposing the resonator in an ambient environment for 168 hours, indicating a significant intra-cavity loss increase from 0.23 dB/cm to 3.1 dB/cm inferred from the coupled mode theory (Fig. 3b). The Q-factor eventually stabilized after two months and settled at $3.5 \times 10^4$ (not shown in Fig. 3b). To ascertain the origin of such loss increase, same measurements were conducted on reference resonator samples stored in a desiccator cabinet and in a high vacuum chamber (pressure $< 10^{-5}$ Torr) after fabrication, and little loss change was observed in both sets of reference samples over time. We therefore conclude that moisture adsorption on the resonator surface accounts for the observed performance degradation. Indeed, the high-Q characteristics can be partially restored by annealing the sample in air at 120 degree Celsius on a hot plate for 30 min. We further note that the post-annealing Q-factor depends on the duration the sample was exposed to the ambient environment: when the sample was exposed for 168 hours, the post-annealing Q-factor was $1.5 \times 10^5$; while samples exposed only for 24 hours can reach a post-annealing Q-factor of $5 \times 10^5$. The observation suggests that other adsorbing species such as hydrocarbon organics possibly also contribute to optical absorption, although further experiments are needed to validate the hypothesis.

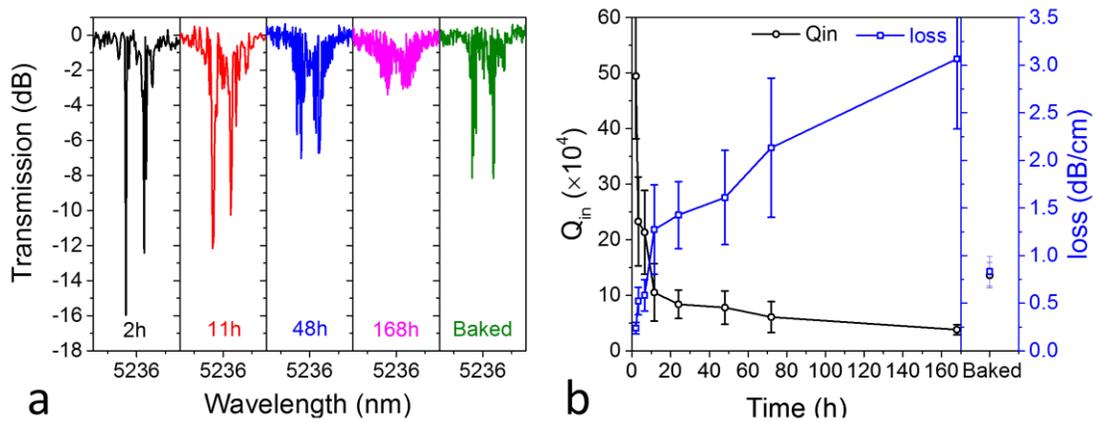

Figure 4. (a) Transmission spectrums of a same resonance peak degraded with time in ambient environment with an average 25% humidity (peak shift due to photosensitivity had been adjust to the position of beginning measurement), and (b) corresponding intrinsic quality factor and propagation loss averaged from three adjacent resonance peaks evolution with time.

We further demonstrated on-chip mid-IR cavity enhanced chemical sensing capitalizing on the resonantly enhanced light-matter interactions in the resonator[15]. In the experiment, cyclohexane solutions containing different concentrations of ethanol was pipetted onto the resonator chip and the resonator transmission spectra were monitored *in-situ*. Cyclohexane was chosen as the solvent as it only exhibit a weak, flat absorption background in the wavelength range of interest (5.1 – 5.3 μm). On the other hand, ethanol has a weak (relative to its main IR absorption band at 3.9 μm which has an absorption coefficient of 2900 cm$^{-1}$) absorption peak centered around 5.2 μm wavelength. Figure 4a shows the resonator transmission spectra as the device was immersed in cyclohexane solutions. The progressive decrease of extinction ratio with increasing ethanol concentration is a consequence of excess optical absorption induced by ethanol. Addition optical absorption induced by ethanol was obtained by subtracting the waveguide loss in pure cyclohexane from the total propagation loss when ethanol was present, and normalized by the modal confinement factor in the sensing region shown as the shaded area in Fig. 2e. The confinement factor is 0.10 based on our finite differential modal simulation for the waveguide we used in the experiments. Concentration-dependent optical absorption of ethanol at 5.2 μm wavelength was plotted in Fib. 4b. The absorption coefficient of ethanol in cyclohexane can be obtained by a linear fit of the plot to be $\alpha_{ethanol}$ = (74 ± 3.4) cm$^{-1}$, which agrees well with measurement result ($\alpha_{ethanol}$ = 78 cm$^{-1}$) obtained on a bench-top Fourier Transform Infrared (FTIR) spectrophotometer.

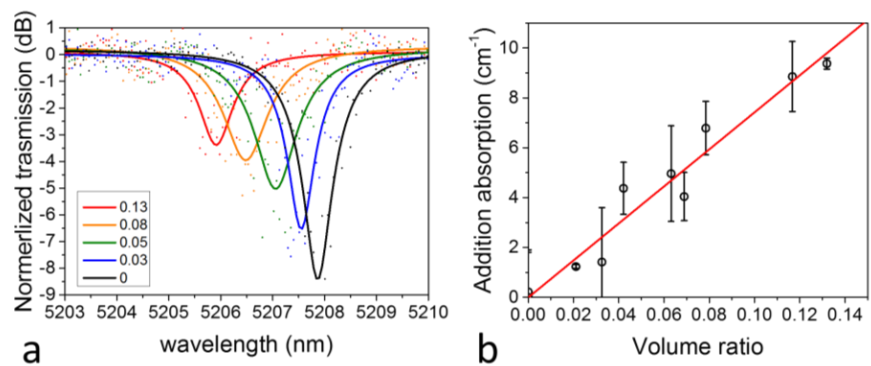

Fig. 5 (a) Mid-IR transmission spectrum of a 200 μm radius disk resonator in ethanol/cyclohexane solutions of different volume ratio (dots are measured transmission intensity, while line is peak fitting by lorentzian function). With increase of volume ratio, the quality factor and extinction ratio decrease along with resonance wavelength shift, (b) addition absorption induced by ethanol was linear fit as functions of ethanol volume ratio.

Conclusion:

In summary, mid-IR $Ge_{23}Sb_7S_{70}$ microdisk resonators on $CaF_2$ substrate were fabricated and tested around 5.2 μm wavelength. The resonators exhibit a quality factor as high as $7 \times 10^5$ (corresponding to a propagation loss of 0.15 dB/cm), which represent the highest value ever reported. Device performance degradation due to moisture was observed and can be partially remedied by baking. We demonstrated mid-IR cavity enhanced chemical sensing by quantifying mid-IR optical absorption of ethanol in cyclohexane, and the result agrees well with FTIR measurement.

Method

Mid-IR $Ge_{23}Sb_7S_{70}$ on $CaF_2$ microdisk resonators were fabricated by lift-off. Firstly, negative photoresist (NR9-3000PY, futurrex, Inc.) was spin coated onto 1 mm thick $CaF_2$ substrates (MTI Corporation) followed by definition of a reversed pattern using a mask aligner (ABM, Inc.). $Ge_{23}Sb_7S_{70}$ glass films of 1.3 μm thickness were then deposited by thermal evaporation at a base pressure lower then $10^{-6}$ torr. The undesired parts of glass film were lift-off by dissolving the resist underneath in acetone, leaving the device pattern on $CaF_2$ substrate. Finally, the devices were cleaved to form facets for optical measurement.

The organic chemicals used in the sensing tests, cyclohexane and ethanol (> 99.5%), were purchased from Sigma-Aldrich. Cyclohexane exhibits a low, broad optical absorption of approximately 20 $cm^{-1}$ near 5.2 μm and was used as the blank solvent. Ethanol has a weak absorption peak (peak absorption about 100 $cm^{-1}$) at 5.2 μm wavelength, and was used as the solute. Their mixtures were prepared based on volume ratios. During the test, the mid-IR resonators were immersed in drop-casted solutions. For the Fourier transformed infrared measurement, solutions were filled into a demountable liquid cell (PIKE Technologies, Inc. 162-1100) with an optical path length varying from 0.1 mm to 0.5 mm and the transmission spectra were recorded by a FTIR spectrometer (Nicolet Magna 860 FTIR).

Supporting Information

Due to the high Q-factors and hence narrow linewidths of the resonance peaks, we introduced a second method to extract the resonator Q-factors and validate the direct spectrum fitting results. In the method, the transmission spectra of the same resonator were recorded as the device was exposed to the ambient environment. The transmission characteristics of the devices can be described using the coupled mode theory, which takes into account coupling between counter propagating resonant modes. As discussed in the main text, water adsorption led to increased optical loss and decrease of both the cavity Q-factor and the extinction ratio. The coupling coefficient κ from waveguide to resonator as well as the coupling strength between the counter propagating modes characterized by the lumped reflectivity R [Ref], however, remained unchanged, as was evidenced by the observation that the peak splitting of the same resonator stayed constant throughout the experiment. When κ and R are fixed, the coupled mode theory specifies a one-to-one correspondence between the intrinsic cavity Q-factor and the extinction ratio, where one parameter can be solved from the other. While direct Q-factor fitting becomes less accurate due to the small resonance peak linewidth, the peak extinction ratio, determined from the minimum transmitted intensity, can be measured reliably through multi-scan averaging. This correlation between Q-factor and extinction ratio allows us to cross-check our fitting results by calculating Q's from measured extinction ratios.